\RequirePackage{ifpdf}
\ifpdf 
\documentclass[pdftex]{sigma}
\else
\documentclass{sigma}
\fi

\begin{document}
\allowdisplaybreaks

\renewcommand{\PaperNumber}{037}

\FirstPageHeading

\ShortArticleName{Regular and Chaotic Regimes in Scalar Field
Cosmology}

\ArticleName{Regular and Chaotic Regimes\\ in Scalar Field
Cosmology}

\Author{Alexey V. TOPORENSKY}

\AuthorNameForHeading{A.V.~Toporensky}

\Address{Sternberg Astronomical Institute, Moscow University,
Moscow,  119899 Russia}
\Email{\href{mailto:lesha@sai.msu.ru}{lesha@sai.msu.ru}}

\ArticleDates{Received October 25, 2005, in f\/inal form March 10,
2006; Published online March 21, 2006}

\Abstract{A transient chaos in a closed FRW cosmological model
with a scalar f\/ield is studied. We describe two dif\/ferent
chaotic regimes and show that the type of chaos in this model
depends on the scalar f\/ield potential. We have found also that
for suf\/f\/iciently steep potentials or for potentials with large
cosmological constant the chaotic behavior disappears.}


\Classification{37B25; 83C75}

\section{Introduction}
The phenomenon of a transient chaos (when a dynamical system
behaves chaotically during a~transient period before reaching some
other regime) becomes in last two decade a matter of intense
investigations. More than 20 years ago it was remarked that a
periodic attractor may follow  a temporal chaotic behavior (see
\cite{kantz} and references therein). After, this kind of chaos
has been found in a broader class of dynamical systems which have
no attractors at all.  In the paper of Gaspard and Rice
\cite{gaspard} the dynamics generated by scattering of a small
disk particle on three hard discs in a plane was described. For an
arbitrary initial data apart from a zero measure
 set
this  particle leaves the system after some time interval (which
can be arbitrarily large), during which the small disc has a chain
of scattering on the three discs.

Several years earlier this kind of behavior was discovered in a
cosmological model describing the evolution of a closed
Friedman--Robertson--Walker (FRW) Universe f\/illed with a massive
scalar f\/ield \cite{Page}. The analog of scattering on a disc in
this model is ``bounce'' -- a transition from a cosmological
collapse to a cosmological expansion of the Universe. The f\/inal
regime of the dynamics for almost all initial condition is
falling into a cosmological singularity. The description of the
set of periodic trajectories in this model in the language of
symbolic dynamics and calculation of topological entropy have been
done in \cite{Cornish}. Later, such analysis was done for
cosmological models with other types of a scalar f\/ield
\cite{we2,te,new}. It appears that depending on the particular
form of the scalar f\/ield potential, the dynamics may be either
chaotic or regular. Several types of transition from chaos to a
non-chaotic dynamics for particular one-parameter families of
potential were described in \cite{new}. In the present paper we
summarize our knowledge on possible regimes in closed FRW
cosmology with a scalar f\/ield.

\section{Equations of motion and basic properties}
We study the following ODE system (the derivation see in
\cite{new}):
\begin{gather}\label{eq1}
\frac{m_{P}^{2}}{16 \pi}\left(\ddot{a} + \frac{\dot{a}^{2}}{2 a} +
\frac{1}{2 a} \right) +\frac{a \dot{\varphi}^{2}}{8} -\frac{a
V(\varphi)}{4} = 0,
\\
\ddot{\varphi} + \frac{3 \dot{\varphi} \dot{a}}{a} + V'(\varphi) =
0.\label{eq2}
\end{gather}
with two variables  -- a scale factor $a$ and a scalar f\/ield
$\varphi$. Here $m_P$ is a constant f\/ixed parameter -- the
Planck mass, the scalar f\/ield potential $V(\phi)$ is a smooth
nonnegative function with $V(0)=0$. The ratio $H \equiv \dot a/a$
is called the Hubble parameter.

This system has one f\/irst integral of motion
\begin{gather}\label{eq3}
-\frac{3}{8\pi} m_P^2 a \left(\dot a^2 +1\right) + \frac{a^3}{2}\left(\dot \varphi^2 + 2V(\varphi)\right)=C.
\end{gather}

This integral play the role of energy and is equal to zero for
cosmological solutions of the system \eqref{eq1}, \eqref{eq2}.
This property arises from the known fact that in General
Relativity the Hamiltonian vanishes identically. That is why in
this section we restrict ourself by $C=0$ three-dimensional
hypersurface in the whole 4-dimensional phase space of
\eqref{eq1}, \eqref{eq2} leaving the general situation to the
Section~4.

The equation \eqref{eq2} for the scalar f\/ield looks like the
equation for a harmonic oscillator with a~time dependent
``friction'' $3 H$ which is positive in an expanding Universe. One
of the most important in modern cosmology, the slow-roll regime
can occurs when this ``friction'' is much larger than the
frequency of the oscillator. More precisely, slow-roll
approximation is characterized by the system
\[
H=\sqrt{\frac{8 \pi}{3 m_P^2} V(\varphi)}, \qquad \dot \varphi=\frac{V'(\varphi)}{3 H},
\]
resulting by neglecting second derivative terms,  kinetic energy
of the scalar f\/ield and spatial curvature. This regime in rather
natural for physically admissible initial conditions \cite{B-Kh,B-Kh1,B-Kh2}
and leads to fast growth of the scale factor $a$ while the scalar
f\/ield $\varphi$ slow rolls toward zero. When the scalar f\/ield
$\varphi$ falls below some value $\varphi \sim m_P$ this regime
disappears~\cite{Linde}.

In the opposite case, when the ``friction'' is small, the dynamics
of $\varphi$ is damping oscillations~\cite{turner}. This regime is
typical for late time evolution of the Universe. However, in
contrast to zero- or negative spatial curvature cases, for a
Universe with a positive spatial curvature this regime is not its
f\/inite fate, and is ultimately followed by a recollapse of the
Universe.

Unlike a recollapse, a transition from contraction to expansion
(often called ``bounce'') being also possible, requires a
specially imposed initial conditions \cite{Star}. These two
characteristic features of a positive spatial curvature case --
ultimate recollapse and existence of initial conditions, leading
to a bounce -- result in a complicated dynamics which in some
situations may be chaotic.

It should be noted that the correct def\/inition of chaos in
General Relativity requires coordinate independent measures of
chaos. For example, Lyapunov exponents do not satisfy this
conditions because they can be set to zero by a simple coordinate
transformations. One of possible invariants, the topological
entropy measures how the number of periodic trajectories grows
with respect to their complexity (see the def\/inition in
\cite{Cornish}). A non-zero topological entropy indicates the
presence of chaos. Its calculation requires a detailed description
of the set of periodical trajectories. A~simpler possibility
appears in transient chaotic system when there are at least two
outcomes following a possible chaotic behavior. In this case we
can plot basins of attraction in the initial condition space. A
fractal nature of boundary between two basins indicates that
dynamics is chaotic, while non-chaotic dynamics leads to smooth,
regular boundaries (see~\cite{basin} for a detailed description of
this method).

\newpage

\section{The nature of chaotic regime}

A very important property of the constraint equation \eqref{eq3}
is that $\dot a^2$ and $\dot \varphi^2$ enter in the left-hand
side with opposite signs. Rewriting \eqref{eq3} in the form
\begin{gather}\label{eq4}
-\frac{3m_{P}^2}{8 \pi}  \frac{\dot{a}^{2}}{a^2}
+\frac{\dot{\varphi}^{2}}{2} = \frac{3m_{P}^2}{8 \pi}
\frac{1}{a^2} -  V(\varphi)
\end{gather}
it is easy to see that
\begin{itemize}\vspace{-2mm}\itemsep=0pt
\item There are no forbidden regions in the conf\/iguration space
$(a, \varphi)$. \item The conf\/iguration space is divided into
two regions -- the region where the right-hand side of~\eqref{eq4}
is positive (and possible extrema of the scale factor are
located), and the region where it is negative (where possible
extrema of the scalar f\/ield are located).\vspace{-2mm}
\end{itemize}

The boundary between these two regions is the curve
\begin{gather}\label{eq5}
a^2=\frac{3}{8 \pi}\frac{m_{P}^2}{V(\varphi)}.
\end{gather}
Zero-velocity points  ($\dot a=\dot \varphi=0$) of a trajectory,
if they exist, should lie on this curve.

Consider points of maximal expansion and those of minimal
contraction, i.e.\ the points, where $\dot{a} = 0$ in more detail.
They can exist only in the region where
\begin{gather}\label{eq6}
a^{2} \leq \frac{3} {8 \pi}  \frac{m_{P}^2}{V(\varphi)} ,
\end{gather}
Using \eqref{eq1}, we can show that  possible points of maximal
expansion ($\dot a=0$, $\ddot a<0$) are localized inside the
region
\begin{gather}\label{eq7}
a^{2} \leq \frac{1}{4 \pi} \frac{m_{P}^{2}}{V(\varphi)}
\end{gather}
while possible points of minimal contraction ($\dot a=0$, $\ddot
a>0$) lie outside this region \eqref{eq7} being at the same time
inside the region \eqref{eq6} (see Fig.~1).

\begin{figure}
\centerline{\includegraphics[width=6cm]{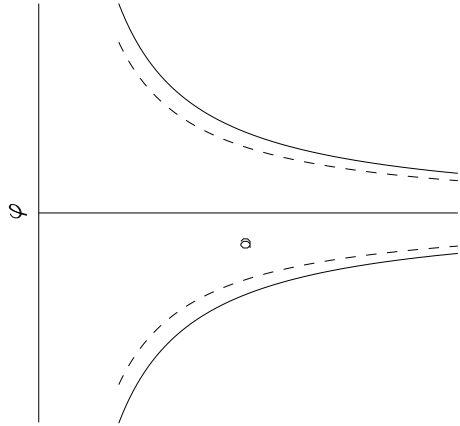}}
\caption{The boundary \eqref{eq5} (solid curves) for the potential
$V=m^2\varphi^2/2$. Points of maximal expansion are located
between two dashed curves while bounces are possible in two narrow
zones between solid and dashed curves.}
\end{figure}

In  \cite{we} it was shown that the region of possible points of
maximal expansion has quite a~regular structure. In the part close
to axis $a = 0$  of this region there are points of maximal
expansion, such that a trajectory, starting from them after some
f\/inite time reaches a point $a=0$, $\varphi=\infty$. In this
point the equations of motion become singular and the trajectory
can not be continued further in time. In General Relativity, this
situation is called as a cosmological singularity. For bigger
initial values of the initial scale factor one can see a narrow
region where a trajectory experiences a bounce i.e.\ goes through
the point of minimal contraction, after that the scale factor $a$
along the trajectory begins to grow. Increasing initial $a$
further, we obtain  the region where a trajectory has a
``$\varphi$-turn'' i.e.\ has the extremum of the scalar f\/ield
$\varphi$ and then falls into singularity. Then one has a region
corresponding to trajectories having bounce after one oscillation
in $\varphi$ and so on.

To avoid a misunderstanding, let us indicate once more what
initial condition space we use. When we start from maximal
expansion point, we f\/ix one time derivative ($\dot a=0$). So, to
f\/ix the initial condition completely, we need only to specify
initial values of $a$ and $\varphi$. The initial~$\dot \varphi$ is
determined from the constraint equation \eqref{eq3}, and our
initial condition space becomes ($a,\varphi$). We will study the
structure of this space, investigating the location of points of
maximal expansion, starting from which a trajectory has a bounce,
but not the location of bounce itself.  The $\varphi=0$
cross-section of regions, leading to bounce, are called as
``bounce intervals''.

Remembering that a trajectory describing an expanding universe
must have a point of maximal expansion, we can apply this analysis
further to bouncing trajectories after a bounce. Their second
point of maximal expansion may lie either inside bounce regions,
or between them. This fact generates a substructure of the region
under consideration. The structure of subregions leading to two
bounces repeats in general the structure of regions having at
least one bounce, and so on and so forth. Continuing this process
{\it ad infinitum}, we get the fractal zero-measure set of
inf\/initely bouncing trajectories escaping the singularity.

If we distinguish two possible singular outcomes, $\varphi \to +
\infty$ and $\varphi \to - \infty$, the method of basins
boundaries can also be applied. Suppose that initial velocity in
the point of maximal expansion is directed ``up'' (initial $\dot
\varphi >0$). Then trajectories with even number of
$\varphi$-turns will approach $\varphi \to + \infty$ singular
point, while trajectories with odd number of $\varphi$-turns fall
into $\varphi \to - \infty$ singularity. A boundary between these
two basins is fractal, indicating the presence of chaos (see
numerical examples in \cite{Cornish}).

In the f\/irst paper on chaos in FRW cosmology \cite{Page} D. Page
have used another description. Instead of starting from a maximum
expansion point, he began the analysis from zero-velocity point at
the curve \eqref{eq5}. Depending on the initial point, two
dif\/ferent situations can be distinguished. Trajectories, going
from this curve into the region \eqref{eq6} (the space between two
solid hyperbolae in Fig.~1) have a point of maximal expansion soon
after start, and then go towards a~singularity.
 So, to prevent an almost immediate collapse, a trajectory
must be directed into the region, where extrema of the scalar
f\/ield are located (outside a solid hyperbola in Fig.~1). These
two situations are separated by a particular trajectory, tangent
to the curve \eqref{eq5}. It means that this trajectory has
\[
\frac{\ddot{\varphi}}{\ddot{a}} = \frac{d \varphi}{da}
\]
at the initial point. Here $\varphi(a)$ in the right-hand side is
the equation of the curve \eqref{eq5}. This point was f\/irst
introduced by Page in \cite{Page} for massive scalar f\/ield
potential $V(\phi)=m^2 \phi^2/2$. In this case
 \begin{gather*}
 \varphi_{\rm page} = \sqrt{\frac{3}{4\pi}} m_{P},\qquad
a_{\rm page}=1/m ,
\end{gather*}
 except for the
trivial solution $\varphi=0$, $a= \infty$.

If  at the initial point on the boundary \eqref{eq5}
\begin{gather}\label{eq9}
\frac{\ddot \phi}{\ddot a} < \frac{d\varphi}{da},
\end{gather}
so that a trajectory is directed outside of the region
\eqref{eq6}, the Universe experiences a long enough expansion
phase. For a massive scalar f\/ield \eqref{eq9} is satisf\/ied if
$\varphi>(3/4\pi)m_{P}$. It should be noted, that {\it a priori}
the signif\/icance of such trajectories for the chaotic structure
is not evident, because we study now only those with a
zero-velocity point. However, numerical data show that a large
class of chaotic trajectories has zero-velocity points; in
particular, all primary periodical orbits (i.e.\ having one bounce
per period) contain such points (see examples in \cite{Cornish}).

With use of  the equation of motion \eqref{eq1}, \eqref{eq2} the
criterion \eqref{eq9} gives for an arbitrary scalar f\/ield
potential
  \begin{gather}\label{eq10}
 V(\varphi)>\sqrt{\frac{3
m_{P}^2}{16 \pi}}V'(\varphi).
 \end{gather}

The condition \eqref{eq10} may be treated as a restriction of
local steepness of the function $V(\varphi)$. In can be easily
seen that for power-law potentials there exists a  value
$\varphi_{\rm page}$ such that for all $\varphi > \varphi_{\rm page}$
the inequality \eqref{eq10} is satisf\/ied. For steeper potentials
the situation changes. For example, the potential
$V(\varphi)=M_{0}^4(\cosh(\varphi/\varphi_0)-1)$ has a Page point
only if $\varphi_0 > \frac{\sqrt{3}}{4 \sqrt{\pi}} m_P$. In the
opposite case the condition~\eqref{eq10} is never satisf\/ied and
all trajectories starting from the curve~\eqref{eq5} go into the
region \eqref{eq6}, experience the point of maximal expansion and
fall into a singularity. Periodical trajectories with a
zero-velocity point can not exist. It has been conf\/irmed
numerically that the chaos is absent in this case (periodical
trajectories disappear, and the boundary between basins of
attractions of $\varphi \to + \infty$ and $\varphi \to - \infty$
singularities becomes smooth) \cite{new}. Although bounces with
$\dot a=0$, $\ddot a>0$ are still possible, trajectories with
bounces have a zigzag form, do not exit the region~\eqref{eq6},
and restore their direction to a singularity soon after the bounce
(see Fig.~2).

\begin{figure}
\centerline{\includegraphics[width=6.3cm]{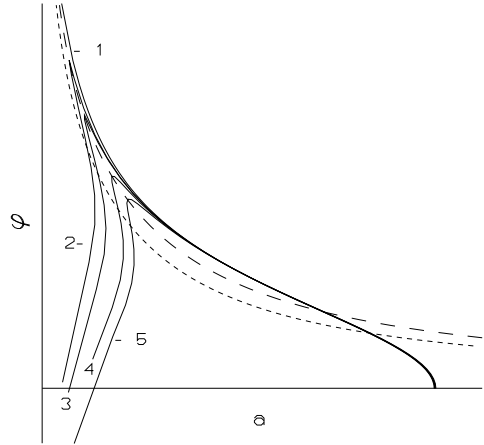}}
\caption{Example of trajectories with the initial conditions close
to the boundary separating trajectories falling into $\varphi=+
\infty$ (trajectory $1$) and $\varphi=-\infty$ (trajectories
$2$--$5$) singularities for the case $\varphi_0 <
\frac{\sqrt{3}}{4 \sqrt{\pi}} m_P$. This boundary is sharp, no
fractal structure is present.  Trajectories $2$--$5$ have a
zigzag-like form, no periodical trajectories are present. The
long-dashed line is the boundary \eqref{eq5}, the short-dashed
line separates zones of bounces from zones of maximal expansion
points.}
\end{figure}

For  steeper potentials like
\[
 V(\varphi)=
M_{0}^4
\big(\exp\big(\varphi^2/\varphi_{0}^2\big)+\exp\big(-\varphi^2/\varphi_{0}^2\big)-2\big)
\]
the condition \eqref{eq10} is def\/initely violated for large
$\varphi$ but, depending on $\varphi_0$ it can be satisf\/ied for
intermediate $\varphi$. If so, zero-velocity points of periodical
trajectories are located between two Page points. For the
potential written above, Page points exist for $\varphi_0 > 0.905
m_{P}$. Our numerical simulations show that the chaotic behavior
exists if $\varphi_0>0.96 m_{P}$. So, we have a rather accurate
and easily calculable condition for existence of the chaotic
dynamics in the system \eqref{eq1}, \eqref{eq2} (another example
of a very steep potential see in~\cite{new}).

\section[Non-zero $C$]{Non-zero $\boldsymbol{C}$}
Now we return to general properties of \eqref{eq1}, \eqref{eq2}
with an arbitrary value of the energy integral~$C$ in~\eqref{eq3}.
One particular case of this problem have already been
investigated. Namely, it can be shown that our dynamical system
with {\it positive} $C$ if formally equivalent to the system
describing a scalar f\/ield in the presence of a pressure-less
matter. A detailed study of this case for massive scalar f\/ield
potential $V(\phi)=m^2 \varphi^2/2$ have been done in~\cite{te}.
The structure of the periodical trajectories becomes more
complicated in comparison with $C=0$ case. First of all, the
number of bounce interval becomes f\/inite and diminishes with
increasing $C$. Besides, we have a nontrivial set of selection
rules, which should be satisf\/ied for any allowed sequence of
intervals along an arbitrary trajectory (in general, the bigger is
the ordinal number of the f\/irst interval, the smaller should be
a maximum possible number of the second one). One particular
example of such rules (simplif\/ied a little in comparison with
those seen in computer simulations) was described in \cite{te}
with calculation of corresponding topological entropy. If $C$
increases, the topological entropy decreases (the width of initial
space regions leading to bounce decreases also), and for $C m >
0.023 m_{P}^2$ the chaos disappears.

For positive $C$ this situation is general -- chaos disappears for
energy integral exceeding some critical value which depends on
particular form of the potential. A negative $C$ alter the nature
of chaos in a dif\/ferent way. Though there are no direct physical
applications of a system~\eqref{eq1}, \eqref{eq2} with a negative
$C$, mathematical properties of this case are very interesting.
Besides, there are dynamical systems of the form similar 
to~\eqref{eq1}, \eqref{eq2} with physically admissible negative
$C$~-- they appear in brane cosmology~\cite{brane}.

When $C$ goes to the range of negative values, the width of the
bounce regions steadily increases.  As an example, we continue to
investigate the massive scalar f\/ield case. 
One of our numerical plots is shown in Fig.~3. We plotted the
$\varphi=0$ cross-section
 of bounce regions depending on~$C$.
This plot represents a~situation, qualitatively dif\/ferent from
studied previously for positive or zero $C$. Namely, the bounce
intervals can merge.

\begin{figure}
\centerline{\includegraphics[width=7cm]{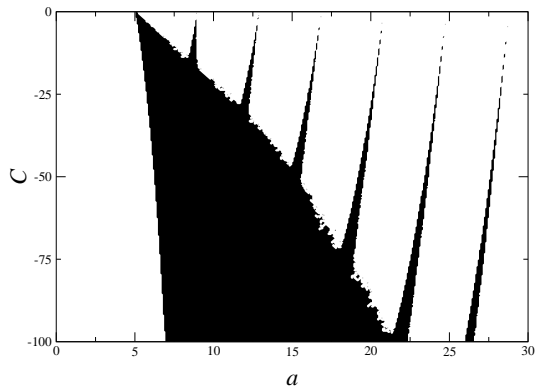}}
\caption{The $\varphi=0$ cross-section of the bounce intervals for
the potential $V=m^2 \varphi^2/2$ and negative~$C$. Consecutive
merging of $5$ f\/irst intervals can be seen in this range
of~$C$.}
\end{figure}

Let us see more precisely what does it mean. For $C>-20$ the
picture is qualitatively the same as for $C=0$ -- trajectories
from 1-st interval have a bounce with no $\varphi$-turns before
it, trajectories which have initial point of maximal expansion
between 1-st and 2-nd intervals fall into a singularity after one
$\varphi$-turn,
 those from
2-nd interval have a bounce after 1 $\varphi$-turn and so on. For
$|C|$ a bit larger than the f\/irst merging value, the 2-nd
interval contains trajectories with 2 $\varphi$-turns before
bounce, the space between 1-st interval (this interval is now a
product of two merged intervals) and the 2-nd interval contains
trajectories falling into a singularity after two $\varphi$-turns.
There are no trajectories falling into a singularity with exactly
one $\varphi$-turn. Trajectories from the 1-st interval can
experience now a complicated chaotic behavior, which can not be
described in a way, similar to the $C=0$ case.

With $C$ decreasing further, the process of interval merging
continues leading to a growing chaotization of trajectories. When
$n$ intervals are merged together, only trajectories with at least
$n$ oscillations of the scalar f\/ield before falling into a
singularity are possible. Those having exactly $n$ $\varphi$-turns
have their initial point of maximal expansion between the 1-st
interval and the 2-nd one (the second interval now contains
trajectories having a bounce after $n$ $\varphi$-turns). For
initial values of the scale factor larger then those from the 2-nd
interval, the regular quasi-periodic structure described above is
restored.  On the other hand, initial values from the 1-st
interval lead to a very complicated structure of chaotic
trajectories. Signif\/icant part of them do not fall into a
singularity for an arbitrary long time of computer simulations.
This fact indicates the presence of a strong chaotic regime, which
is qualitatively more complex than the regime corresponding to a
rather regular fractal structure described in the previous
section. The structure of strong chaos requires further
investigations.

\section{Less steep potentials}

So far we have found that a positive $C$ as well as  steep
potentials are less favorable for chaos. To prove these
dependencies further we consider potentials, less steep than the
quadratic one. We will investigate a common family of potentials
having power-low asymptotic -- Damour--Mukhanov
potentials~\cite{Damour}. They were originally introduced to show
a possibility for the Universe to have an  inf\/lationary
expansion without the slow-roll regime. The explicit form of
Damour--Mukhanov potential is
\begin{gather}\label{eq11}
V(\varphi)=\frac{M_{0}^{4}}{q} \left[ \left(1+\frac{\varphi^2}
{\varphi_{0}^{2}} \right)^{q/2}-1 \right].
\end{gather}
with three parameters $M_0$, $q$ and $\varphi_0$.

For $\varphi \ll \varphi_0$ the potential looks like the massive
one with the ef\/fective mass $m_{\rm eff}=M_{0}^{2}/\varphi_0$.
In the opposite case of large $\varphi$ it grows like $\varphi^q$.

Changing  $M_0$ leads only to rescaling $a$ and  does not alter
the type of chaos. So, we have a two-parameter ($q$ and
$\varphi_0)$ family of potentials with dif\/ferent chaotic
properties. The main result of our numerical studies is that the
event, described in the previous section -- existence of a strong
chaos regime -- takes place also for this family of gently sloping
potentials even with $C=0$.
 Numerical studies show the following picture (see Fig.~4): for
a small enough  $q$ there exists a corresponding critical value of
$\varphi_0$ such that for $\varphi_0$ less than the critical one,
the strong chaotic regime exists. Increasing $q$ corresponds to
decreasing the critical $\varphi_0$.

\begin{figure}
\centerline{\includegraphics[width=9cm]{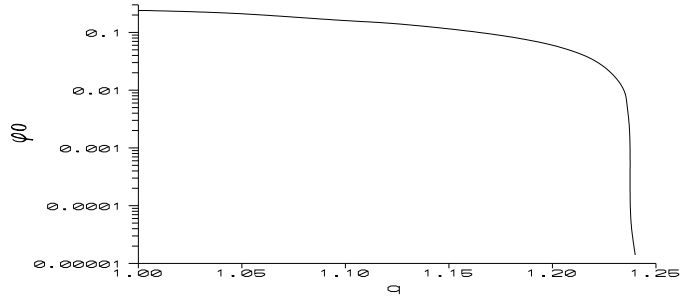}}
\caption{ The value $\varphi_0$ of the potential \eqref{eq11}
corresponding to the f\/irst merging of the bounce intervals
depending on power index $q$.}
\end{figure}

 Since this regime is absent for quadratic and
steeper potentials, $q$ must at least be less than~$2$. We can see
clearly the strong chaotic regime for $q< 1.24$. Near this value
the critical $\varphi_0$ decreases very sharply.

To study further the existence of the strong chaotic regime in
gently sloping potentials we consider another family of potentials
with the same asymptotic for large $\varphi$ and another behavior
for small $\varphi$:
\begin{gather}\label{eq12}
V(\varphi)=\frac{M_{0}^{4}}{q} \left[ \left(1+\frac{\varphi^4}
{\varphi_{0}^{4}} \right)^{q/4}-1 \right].
\end{gather}
For small $\varphi$ the potential $V(\varphi) \sim \lambda
\varphi^4$ with $\lambda=M_0^4/(4 \varphi_0^4)$, though for large
$\varphi$ we still have $V(\varphi) \sim \varphi^q$. Our numerical
studies have shown that the critical $\varphi_0$ in this case is
close to the number obtained for Damour--Mukhanov potentials and
has the same asymptotic value  $ \sim 1.24$. We conclude that
existence of the strong chaos depends mainly on the parameter~$q$.

So far we found that decreasing steepness of the potential acts as
decreasing~$C$. The opposite is also true -- if we start from a
strong chaos for some suitable gently sloping potential with
$C=0$, and then  allow $C$ to increase, the bounce intervals
consecutively separate, strong chaos disappears and with further
increasing $C$ the chaotic regime disappears completely~\cite{PT}.

\section[Potentials, steeper than quadratic one with negative $C$]{Potentials,
steeper than quadratic one with negative $\boldsymbol{C}$}

In the area of potentials steeper than the quadratic one and
negative $C$ we can see two tendencies acting in opposite
directions. We begin with presentation of our numerical results on
the strong chaotic regime. We already know that the strong chaos
exists for $V=m^2 \varphi^2/2$ with suf\/f\/iciently large
negative $C$. On the other hand, our numerical analysis have shown
that for the potential $V(\varphi) \sim \varphi^4$ there is no
strong chaos for any $C$. In order to f\/ind the biggest power
index allowing the strong chaos regime, we use the potential
family \eqref{eq12}.  Our numerical data lead to conclusion that
large enough negative $C$ could produce the strong chaos for $q <
2.15$.

As for possibility for chaotic regime itself, the value of $C$
alters the critical $\varphi_0$ in the exponentially steep
potentials $V \sim (\cosh(\varphi/\varphi_0)-1)$, studied in the
Section~3. Negative $C$ makes the chaos possible for a wider range
of $\varphi_0$. In the limit $C \to -\infty$ the analog of Page
point equation  does not depend on $C$:
  \begin{gather*}
\sqrt{3} V(\varphi)>\sqrt{\frac{ m_{P}^2}{4 \pi}}V'(\varphi)
 \end{gather*}
and, so, the chaos is absent if $\varphi_0$ is less than
$m_{P}/\sqrt{12 \pi}$ for any value of $C$.

\section {Conclusions}

We have investigated possible regimes for the dynamical system
describing a FRW cosmological model with a scalar f\/ield, and the
generalization of this system to a non-zero ``energy
parame\-ter''~$C$. The qualitative features of this dynamics
depend on the steepness of the scalar f\/ield potential $V(\phi)$
and the value of the ``energy parameter''. The general property is
that the chaotic behavior is more signif\/icant for less steep
potentials and lower values of~$C$. The known description of chaos
in FRW cosmology with a massive scalar f\/ield~\cite{Cornish,we}
 appears to be
typical for power-law potentials. In this case only trajectories
starting from narrow quasi-parallel zones in the initial condition
space $(a, \varphi)$ from the point of maximal expansion do not
fall into a~singularity at the f\/irst contraction stage. The full
description of possible periodic trajectories can be done easily
using a suf\/f\/iciently simple coding~\cite{Cornish}.

For less steep potential, however, these zones can intersect and
merge, and the behavior of trajectories becomes more complex. In
FRW cosmology ($C$=0) this strong chaotic regime can appear if
asymptotically $V(\phi) \sim \phi^p$ with $p < 1.24$, in the
general case corresponding power index can reach $2.15$  if $C$ is
negative and large.

For potentials steeper than power-law the chaotic regime can
disappear. We have studied exponentially steep potential in the
form $V(\phi)=M_0^4(\cosh(\phi/\phi_0)-1)$. The condition for
chaos  is $\phi_0>4 m_{P}/ \sqrt{3}$ for $C=0$. Negative C makes
this critical value of $\phi_0$ lower. For $C \to -\infty$ the
critical $\phi_0$ tends to $m_{P}/2 \sqrt{3 \pi}$, being $1.5$
times less then for the $C=0$ case.

A suf\/f\/iciently large positive $C$ kills chaos for an arbitrary
potential $V(\phi)$.

\appendix
\section{Appendix}

In this appendix we brief\/ly describe several dynamical systems
which are generalization of \eqref{eq1}, \eqref{eq2} in
dif\/ferent cosmological scenarios. In all examples we consider
only $C=0$ case.

\subsection{Cosmological constant}

If we ad a constant term $L$
(often written in the form $L=(m_P^2/8\pi)\Lambda$, where $\Lambda$ is
the cosmological constant) to a potential $V(\varphi)$,
the cosmological dynamics changes signif\/icantly. A new late-time
attractor (called the DeSitter regime) with $H \to \sqrt{\Lambda/3}$ (and, correspondingly, $a \sim \exp(Ht)$)
appears. Trajectories, reaching this attractor have no maximal
expansion point. Large enough $\Lambda$ leads to a situation when
all bouncing trajectories falls into the DeSitter regime, and
chaotic behavior disappears~\cite{we2}. For a massive scalar
f\/ield this happens if $\Lambda$ is larger than $\sim 0.3 m^2$.
It should be noticed, that the Page points for the potential
$V(\varphi)=L+m^2\varphi^2/2$ disappear for $\Lambda=0.75
m^2$ \cite{new}, so the Page points criterion does not work well
in this case.

\subsection[Complex scalar field]{Complex scalar f\/ield}

For description of a scalar f\/ield with non-zero charge a
formalism of complex scalar f\/ield is used. The most natural
representation of the complex scalar f\/ield has the form
\begin{gather*}
\phi = x \exp (i \theta),
\end{gather*}
where $x$ is the absolute value of the f\/ield while $\theta$ is
its phase. This phase is a cyclical variable corresponding to the
conserved quantity -- a classical charge $Q \equiv a^3 x^2
\dot\theta$. The equations of motion are \cite{complex}
\begin{gather*}
\frac{m_{P}^{2}}{16 \pi}\left(\ddot{a} + \frac{\dot{a}^{2}}{2 a} +
\frac{1}{2 a}\right) +\frac{a \dot{x}^{2}}{8} -\frac{a V(x)}{4} +
\frac{Q^{2}}{4 a^{5} x^{2}} = 0
\end{gather*}
and
\begin{gather*}
\ddot{x} + \frac{3 \dot{x} \dot{a}}{a} + V'(x) - \frac{2
Q^{2}}{a^{6} x^{3}} =0
\end{gather*}
with the f\/irst integral
\begin{gather*}
-\frac{3}{8 \pi} m_{P}^{2} \big(\dot{a}^{2} + 1\big)
+\frac{a^{2}}{2}\left(\dot{\varphi}^{2} + 2 V(\varphi)\right)
+\frac{Q^2}{2 a^4 x^2}  = 0.
\end{gather*}

The ef\/fect of terms containing the charge $Q$ is similar to the
inf\/luence of positive $C$ -- chaotic behavior disappears for a
suf\/f\/iciently large $|Q|$. For $V=m^2 \varphi^2/2$ it happens
if $|Q| m^2>0.056 m_{P}^2$~\cite{te}

\subsection{Brane Universe}

The equation of motion for a Randall--Sundrum brane Universe
\cite{R-S} in high-energy regime has the form \cite{Binetruy}
\begin{gather*}
\frac{\dot a^2}{a^2}+\frac{1}{a^2}= \frac{\kappa^4}{36}\left(
\frac12\dot\varphi^2+V\right)^2+\frac{C}{a^4},\\
\frac{\ddot a}{a}+\frac{2\dot a^2}{a^2} + \frac{2}{a^2}=
\frac{\kappa^4}{48}\big(4V^2-\dot \varphi^4\big)
+\frac{C}{a^4},\\
\ddot\varphi=-3\frac{\dot a}{a}\dot\varphi-V'(\varphi),
\end{gather*}
where $\kappa^2=8\pi/M_{(5)}^3$, $M_{(5)}$ is a fundamental
5-dimensional Planck mass, $C$ is an integration constant.

In the case of $C=0$ the condition \eqref{eq9} is
\begin{gather*}
\frac{\kappa^4}{36} V(\varphi)^3 >
\left(\frac{dV(\varphi)}{d\varphi}\right)^2.
\end{gather*}

It is easy to see that in this case the Page points exist for
exponential and even steeper than exponential potentials. They
disappear only for potentials in the form of inf\/inite potential
wall. The critical case now is the potential
\begin{gather*}
V(\varphi)=\frac{A}{(\varphi-\varphi_0)^2},
\end{gather*}
and chaos disappears for $A<9 M_{(5)}^6/(4 \pi^2)$ \cite{TTU}.

\subsection {Anisotropic Universe}

The next possibility to generalize equations~\eqref{eq1},
\eqref{eq2}, apart from modifying a scalar f\/ield or ef\/fective
theory of gravity (as have been done in the brane worlds scenario)
is to consider a~broader than FRW class of metrics. If we lift the
assumption of spatial isotropy, the evolution of the Universe will
be described by three dif\/ferent scale factors $a$, $b$ and $c$.
The Einstein equations for a closed Universe with a massive scalar
f\/ield take the form
\begin{gather*}
\frac{(\dot a bc)\dot
{}}{abc}+\frac{1}{2a^2b^2c^2}\big[a^4-(b^2-c^2)^2\big]=
\frac{m^2}{4}\varphi^2,
\\
\frac{(a\dot b
c)\dot{}}{abc}+\frac{1}{2a^2b^2c^2}\big[b^4-(a^2-c^2)^2\big]=
\frac{m^2}{4}\varphi^2,
\\
\frac{(ab\dot
c)\dot{}}{abc}+\frac{1}{2a^2b^2c^2}\big[c^4-(a^2-b^2)^2\big]=
\frac{m^2}{4}\varphi^2,
\\
\ddot\varphi+\left(\frac{\dot a}{a}+\frac{\dot b}{b}+\frac{\dot
c}{c}
\right)\dot\varphi+m^2\varphi=0 
\end{gather*}
with the f\/irst integral
\begin{gather*}
\frac{1}{2} \dot\varphi^2+\frac{m^2}{2}\varphi^2=2\left(\frac{\dot
a}{a}\frac{\dot
 b}{b}+\frac{\dot a}{a}\frac{\dot b}{b}+\frac{\dot b}{b} \frac{\dot
c}{c}\right)
+\frac{1}{a^2}+\frac{1}{b^2}+\frac{1}{c^2}-\frac{a^4+b^4+c^4}{2a^2
b^2 c^2}.
\end{gather*}

This system (in the case of empty Universe, where the scalar
f\/ield is absent) is well known due to chaotic oscillations of
scale factors at the contraction stage, when the overall volume
$v=abc$ decreases monotonically -- so called
Belinsky--Lifshits--Khalatnikov chaos \cite{BKL}. On the other
hand, chaotic oscillations of the volume itself due to presence of
a scalar f\/ield have attracted less attention. It is known that
in general a spatial anisotropy destroys bouncing behavior, and
bounce can survive only for small initial deviation from isotropy.
No anisotropic periodical trajectories have been found
yet~\cite{1999}.

As a f\/inal remark, it should be noticed that modern
modif\/ications of General Relativity, elaborated for quantization
of gravity can modify cosmological dynamics signif\/icantly in
their corresponding semi-classical regimes.  For two interesting
examples we note a new non-singular cosmological attractor in
String Gravity \cite{greek,greek1,greek2} and an ultimate bounce  independent of
initial conditions and a scalar f\/ield potential in Loop Quantum
Cosmology \cite{Param} (for  general descriptions of these
theories, see, for example,
\cite{Polyakov,Bento,Martin1,Martin2}).

\subsection*{Acknowledgments}

Author is grateful to A.Yu.~Kamenshchik, I.M.~Khalatnikov,
S.V.~Savchenko, S.A.~Pavluchenko, S.O.~Alexeyev, V.O.~Ustiansky,
P.V.~Tretyakov and Parampreet Singh for discussions and
collaboration in studies of chaos in a scalar f\/ield cosmology.

\LastPageEnding

\end{document}